\documentstyle[12pt]{article}

\begin{document}
\begin{titlepage}

\begin{flushright}
UAB-FT-328\\
February 1994
\end{flushright}

\vspace{\fill}

\begin{center}
        {\LARGE \bf LOOP REPRESENTATION OF THE PARTITION FUNCTION
\vskip.3cm
OF LATTICE U(1) GAUGE THEORY}
\end{center}

\vspace{\fill}

\begin{center}
       { {\large\bf
        J. M. Aroca }
	\vskip 0.5cm
        Departament de Matem\`atiques, \\
        Universitat Polit\`ecnica de Catalunya, \\
        Escola T\`ecnica Superior d'Enginyers de Telecomunicaci\'o, E-08034 \\
        Barcelona, Spain
        \ \\
\vspace{ 7 mm}
        and
\vspace{ 7 mm}
        \ \\
        {\large\bf  H. Fort}
	\vskip 0.5cm
        Grup de F\'\i sica Te\`orica\\
        and\\
        Institut de F\'\i sica d'Altes Energies\\
        Universitat Aut\`onoma de Barcelona\\
        08193 Bellaterra (Barcelona) Spain}
\end{center}

\vspace{\fill}

\begin{abstract}
We introduce in a natural and straigthforward way the $loop$
(Lagrangian) $representation$ for
the partition function of pure compact lattice QED.
The corresponding classical lattice loop action is proportional to the
quadratic area of the loop world sheets.
We discuss the parallelism between the $loop$ description of this model in
terms of world sheets of loops and
the $topological$ representation of
the Higgs (broken) phase for the non-compact lattice QED
in terms of world sheets of Nielsen-Olesen strings.
\end{abstract}

\end{titlepage}

A unified quantum theory which describes the gauge fields and the
gravitation is one of the main goals pursued by the physicists for long time.
A good candidate for accomplishing this
comprehensive framework is the {\em loop formalism}.
This loop approach was introduced
in the early eighties by Gambini and Trias \cite{gt} as a
general analytical Hamiltonian approach based on the properties
of the group of loops \cite{gt1}.
Some time ago, Ashtekar et al \cite{ahrss} showed how the
quantum gravity can also be formulated in terms of loops.

The Hamiltonian techniques for gauge
theories have been
developed during the last decade and they provide interesting results
for several models \cite{gt2}-\cite{af}. On the other hand
a Lagrangian approach in terms of loops has been elusive, due mainly
to the non-canonical character of the loop algebra. This feature forbids
the possibility of performing a Legendre transformation as
a straightforward way to obtain the Lagrangian from the Hamiltonian.
A Lagrangian loop formulation will give rise to new computation
techniques providing a
a useful complement to the Hamiltonian loop studies.
Recently, Gambini et al \cite{aggs} proposed a tentative classical
action in terms of loop variables for the U(1) gauge
theory.
Shortly afterwards, we proved that the lattice version of this action
is equivalent to Villain form for D=2+1 dimensions but is slightly
different for D=3+1 dimensions \cite{af1}. In fact, this action
written in terms of variables directly attached to spatial loops
seems to fail in describing all the dynamical degrees of freedom
for D=4.

	In this letter we show how the
loops, originally thought up
within the Hamiltonian formalism, can be introduced in a natural and
alternative way in the lattice statistical theory. We follow a
different approach to that of reference \cite{af1}: starting with the
Villain form of the action we obtain in a straigthfoward way a new
representation of the partition
function of 4D lattice U(1) as a sum of the world sheets of
electric loops.
We discuss the conection of this classical action of loops with
the Nambu string action and we point out the parallelism of the
loop representation for our model with the dual representation
of the lattice Abelian Higgs
in $D=4$.

\vspace{3 mm}

Our starting point is the partition function for the Villain form
of the compact U(1) action which is given by

\begin{equation}
Z = \int (d\theta ) \sum_{ n }\exp
(-\frac{\beta}{2}\mid\mid \nabla \theta -2\pi n\mid\mid^2)
\label{}
\end{equation}

We use the notations of the calculus of differential forms on the lattice
of \cite {g}, the same which were already used in reference
\cite{af1} and which are briefly described in an Appendix.
In the above expression, $\theta$ is a real periodic 1-form, that is,
a real number $\theta_l \in [0, 2\pi ]$
defined in each link of the lattice; $\nabla$ is the co-border operator;
$n$ are integer 2-forms, defined
at the lattice plaquettes, and $\mid\mid . \mid\mid^2 = <.,.>$.

If we use the Poisson sumation formula
$\sum_n f(n) = \sum_s \int_{-\infty}^{\infty} d\phi f(\phi) e^{2\pi i\phi s}$
and we integrate the continuum $\phi$ variables we get

\begin{equation}
Z = (2\pi \beta)^{-N_p/2} \int (d\theta ) \sum_{s }\exp
(-\frac{1}{2\beta}<s,s>+i<s,\nabla \theta >)
\label{}
\end{equation}
where $N_p$ in the number of plaquettes of the lattice. We can use
the equality:
$<s,\nabla \theta>=<\partial s,\theta >$ and integrating over $\theta$
we obtain a $\delta (\partial s)$. Then,

\begin{equation}
Z = (2\pi \beta)^{-N_p/2}  \sum_{
                \begin{array} {c} s\\
                              \left(  \partial s = 0 \right)
                \end{array}
                 }\exp
(-\frac{1}{2\beta}<s,s>)
\end{equation}
where $s$ are integer 2-forms.
The condition $\partial s=0$ means that we consider only
closed surfaces made of plaquettes with multiplicity equal to
s$\in Z$. If we consider the intersection of one of such surfaces
with a $t=constant$ plane we get an electric loop. In other
words, we have arrived to an expression of the partition
function for compact electrodynamics in terms of the world
sheets of loops: the $loop$ (Lagrangian) representation.
There are other equivalent representations
which can be obtained from
the Villain form. First, we have the the $dual$
representation \cite{s} obtained
essentially by using the Poisson identity and then performing a
duality transformation. Actually, the loop representation
for the compact U(1) gauge model is reached following this procedure
but stoping before the duality transformation.
Second, for any lattice theory with Abelian
compact variables,
the $topological$ or $BKT$ (for Berezinskii-Kosterlitz-Thouless)
representation \cite{bwpp} via the `$Banks-Kogut-Myerson$'
transformation \cite{bkm}. The $BKT$ expression for the partition
function of a lattice theory with Abelian compact variables is given by

\begin{equation}
Z \propto \sum_{
                \begin{array} {c} *t\\
                              \left(  \partial *t = 0 \right)
                \end{array}
                 }\exp
(-\frac{2\pi^2}{g^2} <*t,\hat{\Delta}*t>)
\end{equation}

i.e. a sum over closed $(D-k-2)$ forms $*t$ attached to
the cells $c_{(D-k-2)}^*$ of the dual lattice representing the
topological objects (one sort of topological excitation for each
compact variable), and where $\hat{\Delta}$
represents the propagator operator. In the case of compact
electrodynamics, $*t\equiv *m$
i.e. the topological objects are monopoles (particles for D=2+1 and
loops for D=3+1) and $\hat{\Delta}\equiv \frac{1}{\Box}$.

\vspace{2 mm}

	Returning to the loop representation of the partition
function (equation (3) ) we can observe that the loop action
is proportional to the $quadratic$ $area$

\begin{equation}
A_2 = \sum_{p \in {\cal S}} n_p^2 = <s,s>,
\end{equation}

i.e. the sum of the squares
of the mul\-ti\-pli\-ci\-ties $s_p$ of pla\-que\-ttes which
constitute the loop's
world sheet { \cal S}. It is interesting to note the similarity
of this action with respect to the Nambu action in the continuum
which is proportional to the area swept out by the bosonic string
in four space-time dimensions.
On the other hand,
we know that in the continuum the classical action of
the first example of a topological string-like soliton, the
Nielsen-Olesen vortex, reduces to the Nambu action in the
strong coupling limit \cite{no}.
The Nielsen-Olesen strings are static solutions of the Higgs
Abelian model. When we compare the
$loop$-representation of lattice compact pure $QED$ and the
$BKT$-representation of lattice $Higgs$ non-compact $QED$ \footnote[1]
{We compare with the non-compact instead of the compact version
because this last, in addition to Nielsen-Olesen strings, also has
Dirac monopoles as topological solutions and then we have to
consider open as much as closed world sheets
(see reference \cite{af2}). }
the analogy between both is patent. That is,
in the case of $Higgs$ non-compact $QED$ model
( a non compact gauge field $A_\mu$
coupled to a scalar field $\Phi=|\Phi|e^{i\phi}$) we have
$*t\equiv *\sigma$, where $*\sigma$ represents a 2-form
which corresponds to the world sheet of the topological objects
namely the Nielsen-Olesen strings
\cite{pwz} and $\hat{\Delta}\equiv \frac{1}{\Box+M^{2}}$ (M is the
the mass acquired by the gauge field due to the Higgs mechanism).
Thus, both models consist in a sum over closed surfaces
which are the world sheets of closed electric
strings (loops) and closed
magnetic strings (closed Nielsen-Olesen vortices) respectively.
The corresponding lattice actions are quadratic in the
world sheet variables in both cases.
Moreover, the creation operator of both loops and N.O. strings
is essentially the same: the Wilson loop operator \cite{gt} \cite{pwz}.

\vspace{1cm}

An interesting question is: are the loops no more than a useful
representation or they have a deeper physical meaning?
For example, one can build the dual superconductor picture of
the confinement in QCD in terms of (chromo) electric field tubes,
that is non-Abelian loops. Does this picture hold for the
electromagnetic interaction?
Lattice QED exhibits a confining-deconfining transition, although
in principle, ordinary continuum QED has only one non-confining phase.
However, there are studies which indicate that also there is
a phase transition for QED in the continuum \cite{mn}. In addition to
the usual weak coupling phase, a strong coupling confining phase
exists above a critical coupling $\alpha_c$. This new phase
could explain a mysterious collection of data from heavy ion
collisions \cite{sch}-\cite{cg}.
The unexpected feature is the observation of positron-electron
resonances with narrow peak energy in the range of 1.4-1.8 MeV.
This suggest the existence of 'electro-mesons' in a strongly coupled
phase of QED. Moreover, a new two-phases model of continuum QED
and a mass formula for taking account of the
positronium spectrum in the strong coupling phase has been regarded
recently \cite{az}.
Thus, in principle, we can speculate about the existence in nature of
abelian electric tubes providing a real support for abelian loops.

	Another point is the fact that now we have a Lagrangian
lattice loop description which allows continuing the exploration
of gauge theories and complementing our previous Hamiltonian
studies \cite{af},\cite{af1}.

\vspace {0.2cm}

Also, we are
interested in extending the loop action in such a way to include
matter fields.  The pointed out parallelism between the $loop$
and the $BKT$ representations
can be used as an heuristic guide in order to accomplish
this task.
This correspondence
turns into an even closer one
when we consider both representations of the same model: the Scalar
compact QED. In a companion letter we exploit this point and
show how to include matter fields into the Lagrangian formalism.
\vspace {0.2cm}

We wish to thank R.Gambini for
useful discussions and comments.

\newpage

\appendix{ {\bf APPENDIX} }

\vspace {0.5cm}

A k-form is a function defined on the k-cells of the lattice
(k=0 sites, k=1 links, k=2 cubes, etc.) over an abelian group
which can be {\bf R}, {\bf Z}, or U(1)={reals module 2$\pi$}.
Integer forms can be considered geometrical objects on the lattice.
For instance, a 1-form is a path and the integer value on a link
is the number of times that the path traverses this link.
$\nabla$ is the co-border operator which
maps k-forms onto (k+1)-forms. It is the gradient operator when acting
on scalar functions (0-forms) and it is the rotational on vector functions
(1-forms). We consider the scalar product of k-forms defined
$<\alpha \mid \beta> = \sum_{c_p}\alpha (c)\beta (c)$ where the sum runs
over the k-cells of the lattice. Under this product the $\nabla$
operator is adjoint to the border operator $\partial$ which maps
k-forms onto (k-1)-forms and which corresponds to minus times the usual
divergence operator. The operator $\Box =\nabla \partial +\partial \nabla$
is called the Laplacian and differs only by a minus sign of the current
Laplacian $ \Delta_\mu \Delta_\mu$.

A useful tool to consider is the duality transformation which maps
biyectively k-forms
over (D-k)-forms. We denote by $*p_{c_{D-k}}$ the dual form of the
$p_{c_k}$ form.
Under duality the border and co-border operators
interchange.

In the loop theory exposed before, physical configurations correspond
to distributions of spatial loops at different times. These are
spatial closed integer 1-forms. Then we consider only forms $c$ which
are 0 for temporal links and closed ($\partial c=0$).
The border and co-border operators can be restricted to the spatial sections
(t=constant) and we denote the spatial Laplacian as $\Delta$.

\end{document}